\documentstyle[preprint,aps,eqsecnum,floats]{revtex}   

\newcommand{\beq}{\begin{equation}}
\newcommand{\eeq}{\end{equation}}
\newcommand{\beqa}{\begin{eqnarray}}
\newcommand{\eeqa}{\end{eqnarray}}

\newcommand{\bsnn}{\mbox{$B \to X_s\,\nu\,\bar\nu$}}
\newcommand{\bxnn}{\mbox{$B \to X\,\nu\,\bar\nu$}}
\newcommand{\bqnn}{\mbox{$B \to X_q\,\nu\,\bar\nu$}}
\newcommand{\bsgnn}{\mbox{$B_s \to \gamma \,\nu\,\bar\nu$}}
\newcommand{\bdgnn}{\mbox{$B_d \to \gamma \,\nu\,\bar\nu$}}

\newcommand{\bctn}{\mbox{$B \to X_c\,\tau\,\bar\nu$}}
\newcommand{\bqtn}{\mbox{$B \to X_q\,\tau\,\bar\nu$}}
\newcommand{\Btn}{\mbox{$B \to \tau\,\bar\nu$}}

\newcommand{\Btt}{\mbox{$B \to \,\tau^+\,\tau^-$}}
\newcommand{\Bttx}{\mbox{$B \to\tau^+\,\tau^-\,(X)$}}
\newcommand{\Bqtt}{\mbox{$B_q \to \,\tau^+\,\tau^-$}}
\newcommand{\Bdtt}{\mbox{$B_d \to \,\tau^+\,\tau^-$}}
\newcommand{\Bstt}{\mbox{$B_s \to \,\tau^+\,\tau^-$}}
\newcommand{\Bdstt}{\mbox{$B_{d,s} \to \tau^+\,\tau^-$}}

\newcommand{\bqtt}{\mbox{$B \to X_q\,\tau^+\,\tau^-$}}
\newcommand{\bxtt}{\mbox{$B \to X\,\tau^+\,\tau^-$}}

\newcommand{\bxmm}{\mbox{$B \to X\,\mu^+\,\mu^-$}}

\newcommand{\BR}{{\rm BR}}
\newcommand{\Em}{\mbox{$E_{\rm miss}$}}  
\newcommand{\Rbs}{\mbox{${\scriptstyle \not \!\!\; R}$}}

\def\npb#1{Nucl.\ Phys.\ {\bf B #1}}
\def\plb#1{Phys.\ Lett.\ {\bf B #1}}
\def\prd#1{Phys.\ Rev.\ {\bf D #1}}

\def\zpc#1{Z.~Phys.\ {\bf C #1}}

\begin{document}

\draft

{\tighten
\preprint{\vbox{\hbox{WIS-96/33/Jul-PH}
                \hbox{CALT-68-2059}
                \hbox{hep-ph/9607473} }}

\title{$B \to\tau^+\,\tau^-\,(X)$ decays: first constraints \\ 
  and phenomenological implications}

\author{Yuval Grossman\,$^{a,1}$, Zoltan Ligeti\,$^{b,2}$ and 
  Enrico Nardi\,$^{a,3}$ }

\footnotetext{
\begin{tabular}{ll}
E-mail addresses: & $^1$ftyuval@weizmann.weizmann.ac.il \\
  & $^2$zoltan@theory.caltech.edu \\ & $^3$ftnardi@wicc.weizmann.ac.il
\end{tabular} }

\address{ \vbox{\vskip 0.truecm}
  $^a$Department of Particle Physics \\
  Weizmann Institute of Science, Rehovot 76100, Israel \\
\vbox{\vskip 0.truecm}
  $^b$California Institute of Technology, Pasadena, CA 91125 }

\maketitle

\begin{abstract}%
The study of \Bttx\ decays can provide us with a better understanding of the
third generation, and can be a useful probe of physics beyond the standard
model.  We present a model independent analysis of these decays.  We
classify new physics that can largely enhance the decay rates and we discuss 
the constraints implied by other processes.  Experimentally, flavor changing
neutral current $B$ decays into final state $\tau$'s are still unconstrained. 
Searches for $B$ decays with large missing energy at LEP provide the first
limits.  We estimate that existing data already imply bounds on the \Bdtt,
\Bstt, and \bxtt\ decay rates at the few percent level.  Although these bounds
are over four orders of magnitude above the standard model predictions, they
provide the first constraints on some leptoquarks, and on some R-parity
violating couplings.

\end{abstract}
} 

\newpage

\section{Introduction}

The standard model (SM) of the strong and electroweak interactions provides an
accurate description of the low energy properties of the known elementary
particles.  However, experimental results involving fermions of the third
generation are far less precise than for the first two generations.  For
example, little is known about decays involving more then one third generation
fermions.  From the theoretical point of view, a better knowledge of the
physics of the third generation could help us understand the hierarchy of 
the fermion masses and mixing angles.  As it is the case in some models of new 
physics \cite{Topcolor}, the third generation 
might even be essentially different from the first two. 

The LEP experiments have provided us with 
several new results on third generation fermions, and in particular on 
the $b$ quark.  One type of
measurement for which the LEP environment has advantages over 
symmetric $B$ factories (like CLEO) or  hadron colliders (like CDF) is the
study of $B$ decay modes that produce a large amount of missing energy due to
neutrinos in the final state. The main background to such analyses is the tail
of the semileptonic decay distribution, so decay modes yielding a harder
missing energy spectrum can be effectively measured or constrained. The excess
of events with large missing energy measured at LEP was interpreted as 
the signature of the decay $B\to X_c\,\tau\,\bar\nu_\tau$ followed 
by $\tau\to\nu\,X\,$ \cite{ALEPH,L3,OPAL}, yielding 
\beq \label{bctnlimit}
  \BR(\bctn) = 2.68 \pm 0.34 \% \,.
\eeq
This is in good agreement with the SM prediction $\BR(\bctn)=2.30\pm0.25\%$
\cite{FLNN}, and constrains certain new physics contributions 
\cite{bctn-mhdm}.  The ALEPH Collaboration  also searched for events with
very large missing energy, $E_{\rm miss}>35\,$GeV \cite{ALEPH}.  The
absence of excess events over the background yielded the 90\% confidence 
level upper limit on the exclusive leptonic decay \Btn \cite{ALEPH}
\beq \label{Btnlimit}
  \BR(\Btn) < 1.8 \times 10^{-3} \,.
\eeq

In a recent paper \cite{bsnn} we showed that the same data also imply a bound
on the flavor changing $\bxnn$ decay rate, and we discussed the resulting
constraints on several possible sources of new physics.  Based of the full
LEP--I data sample, the ALEPH Collaboration has recently announced a
preliminary 90\% confidence level limit on this decay mode \cite{warsaw}
\beq \label{bsnnlimit}
  \BR(\bxnn) < 7.7 \times 10^{-4} \,.
\eeq
This limit is only one order of magnitude above the SM prediction, and 
provides important constraints on several new physics scenarios \cite{bsnn}.

Besides these decay modes, the exclusive \Btt\ and inclusive \bxtt\ decays are
also associated with sizable missing energy due to the neutrinos from the 
$\tau$ decays. Since these processes are presently unconstrained, it is 
interesting to see whether any useful limit can be established by analyzing 
the LEP data on large missing energy events. In the SM, $B$ decays into a pair 
of charged leptons are highly suppressed.  However, certain kinds of new 
physics can enhance these decay rates up to several orders of magnitude above 
the SM predictions.  

In section II we study the \Bdstt\ decays from a theoretical point of view. 
Since we are mainly interested in possible new physics contributions, we
present a model independent analysis.  In section III we estimate the limits 
on the $B\to\tau^+\,\tau^-\,(X)$ branching ratios that could be established 
from the LEP data.  In section IV we discuss the constraints that the limits 
on these decays imply on some new physics models.  Section V contains a 
summary and our conclusions.

\section{Theoretical Analysis}

The most general effective four-fermion interaction involving a $b$ quark, 
a $q=d$ or $s$ quark, and a pair of $\tau^+\,\tau^-$ leptons can be written 
in the form 
\beq \label{lorentz}
{\cal L}^{qb}_{\rm eff} =
 G_F \sum_a C^q_a\, (\bar q\,\Gamma_a\,b)\, (\bar\tau\,\Gamma_a\,\tau) +
 G_F \sum_a {C^q_a}'\, (\bar q\,\Gamma_a\,b)\, (\bar\tau\,\Gamma'_a\,\tau) \,,
\eeq
where $\Gamma_a=\{I,\gamma_5,\gamma^\mu,\gamma^\mu\gamma_5,\sigma^{\mu\nu}\}$,
$\Gamma'_a=\Gamma_a\gamma_5$ and $a=\{S,P,V,A,T\}$. In (\ref{lorentz}) we have
factored out the Fermi constant $G_F$ so that the coefficients $C^q_{a}$ and
${C^q_a}'$ are dimensionless.  While the \bqtt\ decay depends on all ten
coefficients $C^q_a$ and ${C^q_a}'$, fewer operators contribute to the leptonic
\Bqtt\ decay.\footnote{We thank David London for bringing this point to our
attention.}

Let us consider the matrix element $\langle0|\,\bar q\,\Gamma_a\,b\,|B\rangle$.
It can only depend on the four-momentum of the $B$ meson, $p_B^\mu$, and
therefore it must vanish for $\Gamma_T=\sigma_{\mu\nu}$ which is antisymmetric
in the Lorentz indices.  The matrix elements of the parity-even operators
($\Gamma_S=I$ and $\Gamma_V=\gamma^\mu$) also vanish due to the pseudoscalar
nature of the $B$ meson.  Finally, for on-shell leptons, the contribution of
the axial-vector operator $\langle0|\,\bar q\,\gamma^\mu\gamma_5\,b\,|B\rangle
\propto p_B^\mu=p_{\tau^+}^\mu +p_{\tau^-}^\mu$ also vanishes when contracted
with the leptonic vector current $\bar\tau\,\gamma^\mu\,\tau$. Hence, leptonic
$B$ decays are induced only by the following three operators 
\beq \label{operators}  
  C^q_P\, (\bar q\,\gamma_5\,b)\, (\bar\tau\, \gamma_5\, \tau) \,, \qquad
    {C^q_P}'\, (\bar q\,\gamma_5\,b)\, (\bar\tau\, \tau) \,, \qquad 
  C^q_A\, (\bar q\,\gamma^\mu\gamma_5\,b)\, 
    (\bar\tau\, \gamma_\mu\gamma_5\,\tau) \,.
\eeq

Using the PCAC relations 
\beqa \label{PCAC}
\langle0|\, \bar q\,\gamma^\mu\gamma_5\,b\, |B\rangle &=& -i f_B\, p^\mu_B \,,
  \nonumber \\* 
\langle0|\, \bar q\,\gamma_5\,b\, |B\rangle &=&
  i f_B\, {m_B^2\over m_b^2+m_q^2} \simeq  i f_B\,m_B \,,
\eeqa
the most general amplitude for the \Bqtt\ decay reads
\beq \label{amplitude}
{\cal A}_q  = i f_B\, m_B\, G_F\, \left[
  \left(C^q_P + {2m_\tau\over m_B}\, C^q_A \right) (\bar\tau\,\gamma_5\,\tau) 
  + {C^q_P}'\, (\bar\tau\,\tau)\, \right]\,,
\eeq
where, for simplicity, we omitted the subscripts $q=d,s$ for the $B$ mass and 
decay constant.  All three operators in (\ref{operators}) 
appear in the SM. Thus, the general result for the total decay 
rate can be read off from Ref.~\cite{SkKa}
\beq \label{rate}
\Gamma(B_q \to \tau^+\,\tau^-) = {G^2_F\, f_B^2\, m_B^3\over 8\pi}\, 
  \sqrt{1-{4 m^2_\tau\over m_B^2}} \, 
  \left[ \bigg| C^q_P + {2 m_\tau\over m_B}\, C^q_A \bigg|^2 + 
  \bigg(1-{4 m^2_\tau\over m_B^2}\bigg)\, \Big| {C^q_P}'\, \Big|^2 \right] .
\eeq

In the SM, ${C^q_P}'$ and ${C^q_P}$ gets contributions from penguin diagrams
with physical and unphysical neutral scalar exchange, and are suppressed 
as $\sim (m_b/m_W)^2$.  
Then, the dominant contribution to the decay rate
comes from 
\beq \label{CqASM} 
\Big(C^q_A\Big)^{\rm SM} = |V_{tq}^*\,V_{tb}|\, {\alpha_{\rm em}(M_W)\over
  \sqrt8\,\pi\sin^2\theta_W}\, Y\!\left({m^2_t\over m^2_W}\right) ,
\eeq
where, at leading order, $Y(x)=(x/8)\,[(4-x)/(1-x)+3x\ln x/(1-x)^2]$ 
\cite{inami-lim}.  Including the small next-to-leading order correction, 
the SM result for the branching ratio is \cite{BBL}
\beq\label{BttSM}
\BR^{\rm SM}(\Bstt) = 8.9\times10^{-7}\,
  \left[{f_{B_s}\over230\,{\rm MeV}}\right]^2 
  \left[{\overline{m}_t(m_t)\over170\,{\rm GeV}}\right]^{3.12} 
  \left({|V_{ts}|\over0.040}\right)^2 \left({\tau_{B_s}\over 1.6\, 
  {\rm ps}}\right)\,.   
\eeq
Compared to this result, the $\Bdtt$ decay rate has an additional suppression 
of about ${|V_{td}/V_{ts}|^2}\sim10^{-2}$. 
 
In general, new physics that can induce large contributions to the coefficients
in (\ref{rate}) is also likely to enhance the rates for other rare processes. 
The existing experimental limits already imply that in several models, the
$\Bqtt$ decays can only occur at rates far below the sensitivity achievable at
LEP.  Therefore, it is useful to classify the contributions which are already
tightly constrained by other measurements, and the ones which are still
unconstrained (or only weakly constrained).  In doing so, we will avoid
referring to any specific model, but we will use $SU(2)$ gauge invariance to
relate operators contributing to the $\Bqtt$ decays to operators which induce
other transitions. 
In the presence of new physics, operators which are
not manifestly $SU(2)$ invariant can also appear \cite{davidcliff},   
suppressed by inverse powers of some large mass scale.
They will not be considered here, since it 
is unlikely that through such contributions the \Bqtt\ decay rates could get
the few order of magnitude enhancement required in order to be observable at
LEP.

$SU(2)$ invariants can be built out of four SM fermions by combining four
singlets, four doublets, or two singlets and two doublets.  While for 
${\bf 1}^4$ and ${\bf 1}^2\times{\bf 2}^2$ there is only one $SU(2)$ invariant,
two different $SU(2)$ invariants arise from ${\bf 2}^4$.  Taking into account
all inequivalent permutations of the $b$, $q$, and $\tau$ fields, and 
recalling that matrix elements of tensor operators vanish, the following 
operators can contribute to the \Bqtt\ decay
\beqa \label{ffops}
  ({\cal O}_0^{MN})^q &=& 4G_F\, (g_{0}^{MN})^q\, 
  (\bar q_M\, \gamma^\mu\, b_M)\, (\bar\tau_N\, \gamma_\mu\, \tau_N)\,, 
  \nonumber\\* 
({\cal O}_0^{\,\prime\, LL})^q &=& 4G_F\, (g_0^{\,\prime LL})^q\, 
  (\bar\tau_L\, \gamma^\mu\, b_L)\, (\bar q_L\, \gamma_\mu \, \tau_L) \,,
  \nonumber\\ 
({\cal O}_{1/2}^{MN})^q &=& 4G_F\, (g_{1/2}^{MN})^q\, 
  (\bar q_M\, b_N)\, (\bar\tau_N\, \tau_M)\,, \qquad\qquad (M\neq N)
  \nonumber\\* 
({\cal O}_1^{LL})^q &=& 4G_F\, (g_1^{LL})^q\, 
  (\bar q_L\, \gamma^\mu\, b_L)\, (\bar\tau_L\, \gamma_\mu\, \tau_L)\,.
\eeqa
In these equations the subscripts $0$, $1/2$, and $1$ denote the isospin of 
the field bilinears, and $M,N=L,R$ label the fields' chirality.
In terms of $(g_I^{MN})^q$, the $C^q_a$ coefficients in (\ref{rate}), 
which are directly constrained by the experimental data, read  
\beqa \label{relCq}
C^q_P &=& - (g_{1/2}^{LR})^q - (g_{1/2}^{RL})^q \,, \qquad\qquad
  {C^q_P}' = (g_{1/2}^{LR})^q - (g_{1/2}^{RL})^q \,,\nonumber\\*
C^q_A &=& (g_0^{LL})^q + (g_{0}^{RR})^q 
  - (g_0^{LR})^q - (g_0^{RL})^q + (g_0^{\,\prime LL})^q + (g_1^{LL})^q  \,.
\eeqa

\begin{table}[b]
  \begin{tabular}{c|c}
\qquad\qquad Constrained operators  \qquad\qquad\qquad  &  \qquad  Decay mode 
  \qquad\qquad\qquad\qquad\qquad  \\ \hline
\quad\qquad\qquad $(g_{0}^{ML})^q$\,,\quad $(g_{1}^{LL})^q$ 
  \qquad\qquad\qquad\qquad\qquad\qquad  &  \bqnn \qquad\qquad\qquad\qquad  \\ 
\quad\qquad\qquad  $(g_{1/2}^{LR})^d$\,,\quad $(g_0^{\,\prime LL})^d$\,,\quad 
  $(g_{1}^{LL})^d$  \qquad \qquad\qquad\qquad 
  & \Btn \qquad\qquad\qquad\qquad  \\ 
\quad\qquad\qquad $(g_{1/2}^{LR})^q$\,,\quad $(g_0^{\,\prime LL})^q$\,,\quad 
  $(g_{1}^{LL})^q$ \qquad\qquad\qquad\qquad   
  &  \bqtn \qquad\qquad\qquad\qquad  \\ 
  \end{tabular}
\vskip 6pt
\caption{Effective couplings of the operators that contribute  to \Bqtt, 
which are constrained by the limits on the decays 
listed in the  second column.} 
\label{tabdec}
 \end{table}

Some of the heavy states which generate the effective operators in
(\ref{ffops}) appear in non-trivial $SU(2)$ multiplets.  Assuming that the mass
splittings between different members of the same multiplet are not too large,
$SU(2)$ rotations leave the overall coefficients in (\ref{ffops}) invariant to
a good approximation.  This allows us to obtain model independent relations
between contributions  to different
transitions of some of the operators in (\ref{ffops}).  
The operators whose contributions to \Btt\ can be constrained
in this way are listed in Table~\ref{tabdec}.  Operators corresponding to the
effective couplings in the first column are related through $SU(2)$ rotations
to operators which induce the decays $\bqnn$, $\Btn$, and  $\bqtn$, as given in
the second column.  These decay modes provide the strongest constraints on
various new physics contributions to \Btt.  The bounds on $\BR(\bqnn)$
(\ref{bsnnlimit}) and on $\BR(\Btn)$ (\ref{Btnlimit}) imply that the
contributions to \Btt\ proportional to the coefficients in the first two lines
of Table~\ref{tabdec} are much below the present experimental sensitivity.
The coefficients $(g_{1/2}^{LR})^s$ and $(g_0^{\,\prime LL})^s$
in the third line are only weakly constrained by the data on $\BR(\bctn)$
(\ref{bctnlimit}).  This is because the  SM contribution to this decay is
large, and its possible interference with the new physics cannot be neglected.  
Since the sign of the interference is not known, (\ref{bctnlimit}) does not 
yield definite limits on the model independent parameters.
{}Finally, the contributions proportional to
$(g_{0}^{MR})^q$ ($M=L,R$) and to $(g_{1/2}^{RL})^q$ are presently
unconstrained, as they are not related to any existing experimental limit.

\section{Experimental Status}

To date, no dedicated experimental search for  \Bttx\ decays has been 
carried out. In this section we discuss the possibilities of 
searching for these decays in current experiments.

The CLEO Collaboration has established limits on $B_d$ decays into any pair of
charged leptons \cite{CLEO}, including different final state flavors, except
for the $B_d\to\tau^+\,\tau^-$ decay mode.  The reason is that in all but this
case the final state contains a muon or electron with a well-defined energy
that can be easily searched for.  The CDF Collaboration has recently
established strong limits on \BR($B_{d,s}\to\mu^+\mu^-$) \cite{CDFbmm}.  These
limits follow from the absence of muon pairs with invariant mass matching
$m_{B_d}$ or $m_{B_s}$.  Because of these selection criteria, the CDF analysis
does not constrain \Bttx\ followed by $\tau\to\mu$.

The stringent UA1 bound $\BR(B\to X\,\mu^+\,\mu^-)<5\times10^{-5}$ \cite{UA1}
has been used to constrain the product of branching ratios $\BR(\bxtt)\times
[\BR(\tau\to\mu\,\nu\,\bar\nu)]^2$, and thus to infer an indirect limit on
\BR(\bxtt) \cite{davidson}.  However, the UA1 Collaboration searched for muons
pairs with large invariant mass, $3.9\,$GeV$<m_{\mu\mu}<4.4\,$GeV.  Muons from
$\tau$ decays would not have passed this cut, so the limit inferred in
\cite{davidson} does not hold.  

We conclude that the existing data still allow for \Bttx\ branching ratios up
to ${\cal O}(10\%)$.  Therefore, in searching for $B$ decays with a
$\tau^+\tau^-$ pair in the final state, measurements at LEP can be competitive
with other searches, and may even yield the best bounds until asymmetric $B$
factories will start operating.  Unlike at $B$ factories running on the
$\Upsilon(4S)$, both $B_d$ and $B_s$ meson decays can be studied at LEP.  Since
the LEP $b$ hadron sample contains about 40\% $B_d$ and 12\% $B_s$ mesons
\cite{ALEPHbrate}, the limits on $B_s$ decays are weaker than the limits on the
corresponding $B_d$ decays by about a factor of 3.3.  

The absence of $B$ decays associated with very large missing energy, which
yielded the limit on \BR(\Btn) (\ref{Btnlimit}), also constrains \BR(\Btt). 
However, compared to the decay \Btn, the \Btt\ mode yields a much softer
missing energy spectrum, since in this case both neutrinos come from secondary
decays.  In addition, to reject background from semileptonic $b$ and $c$
decays, events with charged leptons in the final state are rejected. 
This weakens the limit by an additional factor of about 65\%, corresponding 
to the hadronic $\tau$ branching fraction. Still,
for sufficiently large \Btt\ branching ratios some events would have been seen
in the large \Em\ region studied by ALEPH \cite{ALEPH}.  To obtain the bound
implied by the absence of such events, we need to evaluate the probability of
\Btt\ decay events to pass the $\Em>35\,$GeV cut, relative to that of \Btn\
decays. We estimated this probability with a Monte Carlo simulation
similar to that in \cite{bsnn}, except that now we   
used  the hadronic invariant mass spectrum
in $\tau$ decays as measured by CLEO.\footnote{We thank Alan Weinstein for 
providing us with this spectrum.} 
We also made some simplifying approximations which are not always 
conservative, and could be avoided in a dedicated experimental analysis. 
{}For example, we neglected the effects of the correlation between the 
direction of the missing momenta from the $\tau^+$ and $\tau^-$ decays. 
Nevertheless, we think that our results give a reasonable estimate
of the limits that can be established by a dedicated experimental 
analysis of the LEP data.  We obtain  the following bounds:
\beqa\label{Bttlimit}
  \BR(\Bdtt) &<& 1.5\% \,, \nonumber\\* 
  \BR(\Bstt) &<& 5.0\% \,.
\eeqa
It is interesting to mention that the
first of these limits is probably close to the bound that CLEO may be able to
obtain using the fully reconstructed $B$ decay sample \cite{KLL}.  As we shall
discuss in the next section, in spite of being over four orders of magnitude
above the SM predictions, the limits (\ref{Bttlimit}) yield the first
constraints on some new physics parameters.  

Neutrinos from the \bxtt\ decay yield a missing energy spectrum which is too
soft to produce any signal in the $E_{\rm miss}>35\,$GeV region.  However, for
large enough branching ratios, events from \bxtt\ would enhance the signal in
the missing energy region used to measure the \bctn\ decay.  Taking into
account that also for \bxtt\ selecting hadronic $\tau$ decays 
weakens  the limit by a factor of about 65\%, we estimate 
that a bound  
\beq \label{bxttlimit}
  \BR(\bxtt) < {\cal O}(5\%) 
\eeq
is within the reach of LEP sensitivity.

Before concluding this section, we mention that the radiative decay
$B\to\gamma\,\nu\,\bar\nu$ is also associated with large missing energy.  The
corresponding branching ratios in the SM have been recently estimated to be of
order $10^{-9}$ for $B_d$ and $10^{-8}$ for $B_s$ \cite{bgnn}.  We can obtain
limits on these decays by assuming a missing energy spectrum for the
$B\to\gamma\,\nu\,\bar\nu$ decay, similar to that in \bxnn\ in the limit 
of zero invariant mass for the  final state hadron system.  Taking into
account that only neutral $B$ mesons can decay into $\gamma\,\nu\,\bar\nu$,
while all $b$ hadrons can decay via the $b\to s\,\nu\,\bar\nu$ transition, we
estimate the bounds 
\beqa\label{bgnnlimit} 
  \BR(\bdgnn) &<& 1 \times 10^{-3} \,, \nonumber\\* 
  \BR(\bsgnn) &<& 3 \times 10^{-3} \,.
\eeqa

Our estimates indicate that the rates for $B\to\tau^+\,\tau^-\,(X)$ decays can
be constrained by the missing energy method only at the few percent level.  By
refining the details of the analysis (for example, by optimizing the \Em\
cuts), dedicated experimental searches could probably establish better limits. 
However, regardless of possible such improvements, it is unlikely that 
the missing energy method could yield much stronger bounds.  
Therefore, it is appropriate to
discuss whether similar (or better) limits can be obtained by different
analyses.  A back-of-an-envelope estimate shows that a limit competitive with
our results could be established at LEP by looking for two charged leptons from
$\tau$ decays, coming from a secondary vertex in the hemisphere opposite to a 
tagged $b$.  
It seems to us that also the inclusive lepton spectrum in $B_d$
decays measured by CLEO \cite{1tag} cannot yield more severe constraints than
those found above.  Finally, the decay mode $B\to\gamma\,\nu\,\bar\nu$ could be
effectively searched for, by looking for the decay photon in
the hemisphere opposite to a tagged $b$. Such a search may yield significantly
better limits than our estimates (\ref{bgnnlimit}).  To what extent some of
these analyses could improve the constraints derived above can only be decided
on the basis of more detailed experimental studies.  

\section{New Physics}

In this section we study the constraints on  new physics 
implied by the limits on \Bttx\  decays.  By comparing (\ref{rate})
with the limits on \BR(\Btt) given in (\ref{Bttlimit}) we obtain the 
following constraint on  the coefficients ${C^q_P}'$, $C^q_P$ and $C^q_A$:
\beqa\label{Cqlimits}
\bigg| C^q_P + \frac23\, C^q_A \bigg|^2 + \frac59\, \Big| {C^q_P}'\, \Big|^2 
 &\lesssim& 2.0 \times 10^{-4}\, \left({190\,{\rm MeV}\over f_B}\right)^2\, 
 \left({1.5 \,{\rm ps}\over \tau_B}\right)\, 
 \left[{\BR(B_q\to\tau^+\tau^-)\over 1.0\times 10^{-2} }\right] \nonumber\\*
&\simeq& \cases{ 3.0 \times 10^{-4} \,; &\quad for $q=d$\,, \cr
                 1.0 \times 10^{-3} \,; &\quad for $q=s$\,. \cr}
\eeqa
In the next two subsections, we give two examples of models in which these
bounds yield the first constraints on some parameters: models with light
leptoquarks, and SUSY without R-parity. First we express in terms of the model
parameters the effective couplings of the operators 
$(g_I^{MN})^q$ in (\ref{ffops}), which arise from integrating out
the new heavy states.  Then,
by inserting the expressions (\ref{relCq}) for the relevant ${C^q_a}$
coefficients into (\ref{Cqlimits}), we derive constraints on various couplings.

\subsection{Leptoquarks}

Leptoquarks (LQ) carry both baryon and lepton number, and hence couple directly
leptons to quarks. A comprehensive analysis of the experimental constraints on
the LQ couplings has been given in \cite{davidson}, and is summarized  in
Table~15 of this reference. As discussed above, the limit on \BR(\bxmm)
\cite{UA1} does not constrain \BR(\bxtt).  Therefore, some of the limits on LQ
Yukawa couplings in \cite{davidson} involving the third generation do not
apply.  

Several different types of LQ are possible, and most of them can induce the
\Bttx\ decays.  However, in many cases LQ also mediate the decays \bxnn\ and
\Btn, which are tightly constrained. Therefore, we will concentrate only on 
those cases where  transitions involving neutrinos are not induced.
This can happen either because
of the particular electric charge of the  LQ (for example $|Q|=4/3$), or when
the LQ couplings to the left-handed lepton doublets vanish.  We adopt here the
notations of \cite{davidson}. Scalar and vector LQ are denoted as $S$ and $V$,
while $SU(2)$ singlets, doublets, and triplets are respectively labeled with a
lower index $0$, $1/2$, and $1$.  The types of  LQ  relevant for
\Bttx\ decays, and for which no strong constraints exist from other 
processes, are
\beq\label{SV}
 \widetilde S_0\,, \qquad S_{1/2}\,, \qquad V_0^\mu\,, \qquad V_{1/2}^\mu\,.
\eeq
The relevant scalar and vector terms in the interaction Lagrangian can be 
found in \cite{davidson}.  Schematically, they read 
\beq
  {\cal L}_{LQ} = \lambda_{ij}^{LQ}\, \ell_i\, q_j\, \phi_{LQ}\,,
\eeq
where $\ell_i$ and $q_j$ denote respectively a lepton and quark fields,
$\phi_{LQ}$ represents one of the LQ in (\ref{SV}), and $i$ and $j$ are
generation indices.  In deriving our constraints, we assume that all the
$\lambda_{ij}^{LQ}$ couplings are real, that only one type of LQ is present 
at a time (for other possibilities see \cite{moreLQ}), and we neglect the 
rotation from the interaction to the
mass basis \cite{Miriam}.  Integrating out the  LQ fields and {}Fiertz
transforming, we obtain the expressions for the   unconstrained coefficients 
of the relevant effective four-fermion operators given in (\ref{ffops})
\beqa \label{LQ}
(g_0^{RR})^q  &=& {(\lambda^{\widetilde S_0}_{R})_{3q}\, 
  (\lambda^{\widetilde S_0}_R)_{33}\over 8G_F\, m^2_{\widetilde S_0}}  \,, \
{(\lambda^{V_0}_{R})_{3q}\, (\lambda^{V_0}_{R})_{33}\over  
  4G_F\, m^2_{V_0}} \,;  \nonumber\\[4pt]
(g_0^{LR})^q  &=& {(\lambda^{S_{1/2}}_{R})_{3q}\, 
  (\lambda^{S_{1/2}}_{R})_{33}\over 8G_F\, m^2_{S_{1/2}}}  \,, \   
{(\lambda^{V_{1/2}}_{R})_{3q}\, (\lambda^{V_{1/2}}_{R})_{33}\over 
  4G_F\, m^2_{V_{1/2}}} \,;  \nonumber\\[4pt]
(g_0^{\,\prime LL})^q &=& {(\lambda^{V_0}_{L})_{3q}\, 
  (\lambda^{V_0}_{L})_{33}\over 4G_F\, m^2_{V_0}} \,;  \nonumber\\[4pt]
(g_{1/2}^{LR})^q &=& {(\lambda^{V_0}_{L})_{3q}\, 
  (\lambda^{V_0}_{R})_{33}\over 2G_F\, m^2_{V_0}} \,;  \qquad
(g_{1/2}^{RL})^q = {(\lambda^{V_0}_{R})_{3q}\, (\lambda^{V_0}_{L})_{33}\over 
  2G_F\, m^2_{V_0}} \,.  
\eeqa

{}For the different products of LQ couplings 
${(\lambda^{LQ}_{M})_{32}\, (\lambda^{LQ}_N)_{33}}$ 
involving the second and third generation
fermions, from (\ref{relCq}) and (\ref{Cqlimits}) we obtain  
the following limits: 
\beqa\label{LQlimits}
{\lambda^{\widetilde S_0}_{R}\, \lambda^{\widetilde S_0}_R}\,, \ 
{\lambda^{S_{1/2}}_{R}\, \lambda^{S_{1/2}}_{R}}
 &<& 4.4 \times 10^{-2} 
  \left({m_{LQ}\over 100\, {\rm GeV}}\right)^2\,, \nonumber \\
{\lambda^{V_0}_{R}\, \lambda^{V_0}_{R}}\,, \
{\lambda^{V_0}_{L}\, \lambda^{V_0}_{L}}\,, \
{\lambda^{V_{1/2}}_{R}\, \lambda^{V_{1/2}}_{R}}
&<& 2.2  \times 10^{-2}
  \left({m_{LQ}\over 100\, {\rm GeV}}\right)^2\,, \nonumber \\
{\lambda^{V_0}_{L}\, \lambda^{V_0}_{R}}\,, \
{\lambda^{V_0}_{R}\, \lambda^{V_0}_{L}}
&<&  5.9 \times 10^{-3}
  \left({m_{LQ}\over 100\, {\rm GeV}}\right)^2\,, 
\eeqa
where the indices (32) and (33) respectively for the first and second coupling
in the products are understood. The bounds on the analogous products
${(\lambda^{LQ}_{M})_{31}\,(\lambda^{LQ}_N)_{33}}$ involving a first generation
($q=d$) quark are about a factor of 1.8 stronger.  From Table~I we see that 
for $q=d$ the products of the LQ couplings in (\ref{LQ}) corresponding to
$(g_0^{\,\prime LL})^q$ and $(g_{1/2}^{LR})^q$ are already tightly constrained
by the limit on \Btn\ decays, while for $q=s$ they are weakly constrained by
the upper bound on \bctn.  For all the other combinations, the bounds obtained
from \Bttx\ are the strongest.

\subsection{SUSY without R-parity}

In SUSY models, it is usually assumed that R-parity is a good symmetry. 
However, this is not necessarily the case, and phenomenologically viable models
have been constructed  where R-parity is not imposed as an exact symmetry
\cite{RSUSY}.  In the absence of R-parity, additional baryon and lepton number
violating terms are allowed in the superpotential.  Some of these terms can
induce a large enhancement for certain rare $B$ decay modes.  Denoting by
$L^i_L$, $Q^i_L$, $\ell^i_R$, and $d^i_R$ respectively the chiral superfields of
the $i$'th generation containing the left-handed lepton and quark doublets,  
the right-handed lepton, and the down-type quark singlet, the R-parity 
violating terms relevant for $B$ decays read
\beq \label{Wrpb}
W_{\Rbs} = \lambda^{}_{ijk}\, L^i_L\, L^j_L\, \bar\ell^k_R + 
\lambda^\prime_{ijk}\, L^i_L\, Q^j_L\, \bar d^k_R \,.
\eeq
The terms in (\ref{Wrpb}) give rise to two types of diagrams that can  
mediate \Bttx.  Exchanging a left- or right-handed squark gives rise to 
effective operators proportional to $\lambda^\prime\,\lambda^\prime$.
Since squark exchange induces also the decay \bxnn, these 
operators are already tightly constrained, and hence irrelevant for the 
present discussion.  
The exchange of left-handed sleptons generates operators proportional to 
$\lambda^\prime\,\lambda$.  These operators do not induce \bxnn, but they 
can still contribute to \Btn\ and to \bctn. 
The corresponding effective couplings are: 
\beq \label{RP}
(g_{1/2}^{LR})^q = {\lambda^\prime_{kq3}\, \lambda^{}_{k33} \over 
  4G_F\, m^2_{\widetilde L_k}} \,, \qquad 
(g_{1/2}^{RL})^q = {\lambda^\prime_{k3q}\, \lambda^{}_{k33} \over 
  4G_F\, m^2_{\widetilde L_k}} \,,
\eeq
where $k=1,2$ due to the antisymmetry in the first two indices of the 
$\lambda$ couplings.  Neglecting possible cancelations between the 
contributions from $\widetilde L_1$ and $\widetilde L_2$ exchange, and 
assuming that one of the two couplings $\lambda^\prime_{k3q}$ and 
$\lambda^\prime_{kq3}$ is dominant, from (\ref{relCq}) and (\ref{Cqlimits}) 
we obtain 
\beq
\lambda_{23k}^\prime\, \lambda^{}_{k33}\,, \ 
\lambda_{32k}^\prime\, \lambda^{}_{k33} < 1.2 \times 10^{-2} 
  \left({m_{\widetilde L_k}\over 100\, {\rm GeV}}\right)^2\,.
\eeq
The bounds on the analogous combinations involving the couplings of the  
$d$ quark are about a factor of 1.8 stronger.  From Table~\ref{tabdec} 
we see that more stringent bounds already exist on the combination 
involving $\lambda^\prime_{k13}$, while for $\lambda^\prime_{k23}$ 
additional weak constraints can be derived from \bctn. 
However, for the products involving the $\lambda^\prime_{k3q}$ 
couplings, these are the strongest limits.  

\section{Summary and conclusions}

To date no experiment has searched for \Bttx\ decays, thus no experimental
bounds exist on the corresponding branching ratios.  $B$ decays into final
states involving $\tau$ leptons can be searched for by looking for the missing
energy associated with the neutrinos from $\tau$ decay.  Such searches 
have been carried out at LEP to measure the \bctn\
decay rate \cite{ALEPH,L3,OPAL} and to set bounds on the \Btn\ and \bsnn\
decays \cite{ALEPH,warsaw}. In this paper we pointed out that similar analyses
can also set bounds on \Bttx\ decays.  We estimated that limits at the few 
percent level [see eqs.~(\ref{Bttlimit}) and (\ref{bxttlimit})] are within 
the reach of the LEP sensitivity.  

The SM predictions for the \Bttx\ branching ratios are of order $10^{-6}$ or
below.  Therefore, the current experimental sensitivity only allows us to
derive constraints on physics beyond the SM.  To identify what kind of new
physics could yield large enhancements, we performed a model independent
analysis of these decays.  While the decay \bxtt\ can depend on all the ten
possible Dirac structures (\ref{lorentz}), \Bdstt\ can be induced only by the 
three operators in (\ref{operators}). 
Thus, for studying new physics, these decay modes
are complementary to one another.  
Much could be learned from the individual rates, and also from their ratio.

Operators which can induce the \Bttx\ decays can be related to operators 
contributing to other processes through $SU(2)$ rotations.
In particular, when the
dominant operators involve left-handed lepton doublets, \Bttx\ can be related
to \bxnn\ and to \Btn.  In these cases the limits on these processes 
\cite{warsaw,ALEPH} provide strong constraints.  
Therefore, only operators involving right-handed $\tau$'s can induce \Bttx\ 
at the level of current experimental sensitivity.  This
restricts the types of new physics models that can be constrained through these
decays.  We gave two examples of such models: light leptoquarks, and SUSY
without R-parity. In both cases, the rather weak limits that we have estimated 
already yield the strongest constraints on some of the model parameters.

In the future, $B$ factories will provide significantly larger samples of $B$
decays.  Hopefully, the \Bttx\ decays will be observed, even if their rates are
as small as predicted by the SM.  Measurements of various $B$ decay rates into
final states involving $\tau$'s (or $\nu_\tau$'s) as well as of other
observables (like the $\tau$ polarization \cite{Hewett}) would provide very
valuable information.  Even if the experimental difficulty of these
measurements will only allow establishing tighter limits on the decay rates,
this will still yield strong constraints on possible physics beyond the SM, 
and might provide us with a better understanding of the third generation.

\acknowledgements

Several discussions on theoretical issues with David London 
are warmly acknowledged.  
We  thank Louis Lyons, Yossi Nir, Alan Weinstein and Mark Wise for
conversations.  ZL is grateful to the Particle Physics Department of the
Weizmann Institute of Science for their hospitality.  ZL was supported in part
by the U.S.\ Dept.\ of Energy under Grant no.\ DE-FG03-92-ER~40701.

{\tighten

}  

\end{document}